\documentclass[doublespacing]{elsart}

\def\gaas{GaAs/Al$_x$Ga$_{1-x}$As\ }

\begin{document}

\begin{frontmatter}

{\bf Version Date:} \today \par

\title{
Quantum transport calculations for quantum cascade laser structures 
}

\author{S.-C. Lee and A. Wacker}

\address{
Institut f\"ur Theoretische Physik, Technische Universit\"at Berlin, 
10623 Berlin, Germany
}

\begin{abstract}
We apply a quantum transport theory based on nonequilibrium Green's functions
to quantum cascade laser (QCL) structures, treating simultaneously the 
transmission through the injector regions and the relaxation due
to scattering in the active region. The quantum kinetic equations are 
solved  self-consistently using self-energies for interface roughness 
and phonon scattering processes within the self-consistent Born 
approximation. In this way, we obtain the current density $J$, 
and the average electronic distribution $f(E)$ at a given energy $E$,
as a function of applied bias.
As a test case, we apply the theory to a \gaas
QCL structure reported in the literature. The theoretical
results reproduce well reported voltage-current (V-I) measurements, 
and also demonstrate a population inversion at a bias that
agrees well with the range of currents and fields at
which lasing is observed.
\end{abstract}

\end{frontmatter}

{\bf Address for correspondence:} S.-C.~Lee, 
Institut f\"ur Theoretische Physik, Technische Universit\"at Berlin, 
10623 Berlin, Germany. Tel: 49 30 31424858, Fax: 49 30 31421130,
email: lee@physik.tu-berlin.de

To be published, Physica E (MSS 10 proceedings)
\newpage
 
\parindent=20pt

Quantum cascade laser (QCL) heterostructures\cite{Fai94a} are unipolar devices
which operate on the basis of light-emitting intersubband (intraband) 
transitions. They are designed with a complicated sequence of semiconductor
layers with varying widths. This sequence of layers is repeated
up to, for example, 30 periods. Part of the semiconductor
layer sequence in each period operates as the active region within 
which the lasing transition occurs. The remaining layers in the period play
the role of an injector for the active region, as well as
a collector of carriers from the preceding period. Fig. 1
shows an example of the conduction band line-up of a QCL structure 
taken from Ref. \cite{Sir98}.

The transport of carriers through QCL structures is governed
by a complex interplay between transmission through the 
injector region, and relaxation in the active region
through scattering processes and light emission
(\cite{Iot01}, \cite{Tor00}). To investigate
the carrier transport in QCL structures, we apply
a quantum transport theory based on nonequilibrium Green's functions 
\cite{Hau96},
which incorporates both transmission in the injector region
and scattering-induced relaxation processes. The quantum
transport equations for the nonequilibrium Green's function
$G^{\rm ret}({\bf k},E)$, and the correlation function $G^<({\bf k},E)$, 
are solved self-consistently with self-energies for phonon and 
interface roughness scattering processes, within the self-consistent
Born approximation. From $G^{\rm ret}({\bf k},E)$ and
 $G^<({\bf k},E)$, we can determine the current density $J$, 
and the average electronic distribution $f(E)$,
which gives the average occupation of the levels at a given energy $E$.

To model a QCL structure we treat the system as a periodic
superlattice structure where each period consists of $N$
barriers and wells with varying widths. 
The Hamiltonian for this structure
may be separated into two parts, $\hat{H} = \hat{H}_o
+ \hat{H}_{\rm scatt} $. The Hamiltonian $\hat{H}_o$ contains the
superlattice potential and, if present, a static electric
field $\mathcal{E}$ applied in the growth direction,
i.e., $\hat{H}_o = \hat{H}_{\rm SL} + \hat{H}_\mathcal{E}$.
A plane-wave basis is used in the plane of the semiconductor layers. 
In the growth direction, we use a localised, 
Wannier state basis \cite{Wac01}.
Hence, we write
\begin{eqnarray}
&&\hat{H}_{\rm SL}\; = \;\sum_{n,\nu}  \Big[ \int d^2k\; 
 E^\nu({\bf k})\, a_{n}^{\nu\dagger}({\bf k})\, a_{n}^\nu({\bf k})
    \nonumber  \\
&&   \qquad\qquad\qquad + \; \int d^2k \;
    T_{1}^\nu \Big(a_{n+1}^{\nu\dagger}({\bf k})\, a_{n}^\nu({\bf k})
    +  a_{n-1}^{\nu\dagger}({\bf k})\, a_{n}^\nu({\bf k})\Big)\Big],
\label{eq.hsl}
\end{eqnarray}   
\noindent where the index $n$ labels a period in the superlattice,
and the index $\nu$ labels a Wannier level $\psi^{\nu}(z)$
within a period.  $a_{n}^{\nu\dagger}({\bf k})$ 
and $a_{n}^\nu({\bf k})$ are creation and annihilation operators for an 
electron with the two-dimensional in-plane wavevector {\bf k},
in the $\nu$th Wannier level, in period $n$. 
The Wannier states are not eigenstates of $\hat{H}_{\rm SL}$, and
therefore $T_{l}^\nu$ represents the off-diagonal coupling between
Wannier levels in different periods, and $E^\nu$ represents the
diagonal elements of $\hat{H}_{\rm SL}$ in this basis. 
We only take into account
couplings between adjacent periods.
The Hamiltonian $\hat{H}_\mathcal{E}$, due to the electric field, is written as
\begin{eqnarray}
&& \hat{H}_\mathcal{E} \; = \; \sum_{n,\nu,\mu} \int d^2k\; \Big[
-e\mathcal{E}R_0^{\mu\nu}a_{n}^{\mu\dagger}({\bf k})\, a_{n}^\nu({\bf k})
\nonumber\\
&& \qquad\qquad\qquad\quad   -e\mathcal{E}R_1^{\mu\nu} 
\Big( \,
a_{n+1}^{\mu\dagger}({\bf k})\, a_{n}^\nu({\bf k}) +
a_{n}^{\nu\dagger}({\bf k})\, a_{n+1}^\mu({\bf k})\,
\Big)
\Big],
\label{eq.he}
\end{eqnarray}
\noindent where $R_l^{\mu\nu} = \int dz\; \psi^\mu(z - ld)\, z \,\psi^\nu(z)$.
$d$ is the length of one period.
The interface roughness and phonon scattering processes 
expressed by $\hat{H}_{\rm scatt}$ are treated in the form of self-energies
which are described below. We have neglected electron-electron
scattering processes in this present work because of the 
increased complexity in the theory when they are included.
These processes should, however, be treated in a more comprehensive study
at a later stage.

The quantum transport equations derived from the Hamiltonian $\hat{H}$
have the following form (\cite{Hau96},\cite{Wac01}). 
The retarded Green's function
$G^{\rm ret}_{\alpha_1 \alpha_2}({\bf k},E)$ 
is given by the
Dyson equation
\begin{equation}
E\, G^{\rm ret}_{\alpha_1\alpha_2}({\bf k},E)  
 - \sum_{\beta}  \Big[(\hat{H}_o)_{\alpha_1\beta} + 
\Sigma^{\rm ret}_{\alpha_1\beta}({\bf k},E) \Big]
G^{\rm ret}_{\beta\alpha_2}({\bf k},E) = \delta_{\alpha_1\alpha_2},
\label{eq.gret}
\end{equation}
where $\alpha_1$ and $\alpha_2$ are general indices that include
both the period and Wannier level indices, e.g. $\alpha_1 \equiv (m,\mu)$.
The correlation function $G^<_{\alpha_1\alpha_2}({\bf k},E)$ is obtained
from the Keldysh relation:
\begin{equation}
G^<_{\alpha_1\alpha_2}({\bf k},E) = \sum_{\beta\beta'} 
G^{\rm ret}_{\alpha_1\beta}({\bf k},E)\, \Sigma^<_{\beta\beta'}({\bf k},E)\, 
G^{\rm adv}_{\beta'\alpha_2}({\bf k},E),
\label{eq.gless}
\end{equation}
where $G^{\rm adv}({\bf k},E) = [G^{\rm ret}({\bf k},E)]^\dagger$.
The self-energies $\Sigma^{\rm ret}({\bf k},E)$ and
$\Sigma^<({\bf k},E)$ describe
the scattering processes arising from $\hat{H}_{\rm scatt}$,
and are evaluated within the self-consistent Born approximation.
Assuming that the diagonal parts of $G({\bf k},E)$ and
$\Sigma({\bf k},E)$ dominate,
the self-energy for interface roughness scattering has the form:
\begin{equation}
\Sigma_{m\mu, m\mu}^{<,{\rm rough}}({\bf k},E) = \sum_{n\nu,{\bf k'}}
\langle |V_{m\mu, n\nu}^{\rm rough}( {\bf k - k'})|^2  \rangle \,
G_{n\nu, n\nu}^<({\bf k'},E). 
\label{eq.sigrough}
\end{equation}
$V_{m\mu, n\nu}^{\rm rough}( {\bf k' - k} )$ is the matrix element
for interface roughness scattering.
The equation for $\Sigma^{\rm ret, rough}$ is obtained by the replacements: 
$\Sigma^{<,{\rm rough}}\, \rightarrow \,\Sigma^{\rm ret, rough}$, 
and $G^{<}\, \rightarrow \, G^{\rm ret}$. 
For optical or acoustic phonon scattering, the self-energies are:
\begin{eqnarray}
 \Sigma_{m\mu, m\mu}^{\rm ret, phon}({\bf k},E)\; &= &\;\sum_{n\nu,{\bf k'}} \, 
|V_{m\mu, n\nu}^{\rm phon}({\bf k},{\bf k'})|^2 \, \nonumber \\
& & \; \times\;\biggr\{
[f_B(E_{\rm phon}) + 1] \,
G_{n\nu, n\nu}^{\rm ret}( {\bf k'},E - E_{\rm phon} ) \nonumber \\
& &\;+\, f_B(E_{\rm phon})\;
G_{n\nu, n\nu}^{\rm ret}( {\bf k'},E + E_{\rm phon} ) \nonumber \\
 & &\;+\,
 \frac{1}{2}\, \Big[ G_{n\nu, n\nu}^<( {\bf k'},E - E_{\rm phon} ) 
-  G_{n\nu, n\nu}^<( {\bf k'},E + E_{\rm phon} ) \Big] \biggl\}
\label{eq.sigret}
\end{eqnarray}
and
\begin{eqnarray}
&& \Sigma_{m\mu, m\mu}^{<, {\rm phon}}({\bf k},E) = \sum_{n\nu,{\bf k'}} \, 
|V_{m\mu, n\nu}^{\rm phon}({\bf k},{\bf k'})|^2 \, \nonumber \\
&& \qquad\qquad\qquad\qquad\quad\biggr\{ f_B(E_{\rm phon})\, 
G_{n\nu, n\nu}^<( {\bf k'},E - E_{\rm phon} ) \nonumber \\
&& \qquad\qquad\qquad\qquad\quad + \, [f_B(E_{\rm phon}) + 1]\, 
G_{n\nu, n\nu}^{\rm ret}( {\bf k'},E + E_{\rm phon} )
\biggl\},
\label{eq.sigless}
\end{eqnarray}
where $V^{\rm phon}$ represents the interaction with optical or 
acoustic phonons, and
$E_{\rm phon}$ represents the energy of the optical or acoustic phonon.
 $f_B(E) = 1/[\exp(E/k_B T) - 1]$ is the equilibrium
phonon distribution at a temperature $T$. 
For practical purposes, we use momentum independent scattering matrix
elements $V_{m\mu, n\nu}^{\rm phon/\atop rough}({\bf k}_{\rm typ},
{\bf k'}_{\rm typ})$, employing a  representative momentum
${\bf k}_{\rm typ}$.
We find the Green's functions $G^{\rm ret}$
and $G^<$ by solving the quantum transport equations, Eqs.~(\ref{eq.gret})
and (\ref{eq.gless}), self-consistently with the equations for the
self-energies, Eqs.~(\ref{eq.sigrough}) -- (\ref{eq.sigless}). 
Periodic boundary conditions are used, i.e. $G_{m\mu, n\nu}({\bf k}, E)
\equiv G_{(m+l)\mu, (n+l)\nu}({\bf k}, E - le\mathcal{E}d)$.

Once $G^{\rm ret}$ and $G^<$ have been determined, 
we can evaluate experimentally-related quantities such as
the current density $J = J^{0 \rightarrow 1}_{\rm coh} +
J_{\rm scatt}^{0 \rightarrow 1}$ between  periods $0$ and $1$ with
\begin{equation}
J_{\rm coh}^{0 \rightarrow 1} = \frac{2e\rho_o}{\hbar}\int_{0}^\infty  dE_k
\int \frac{dE}{2\pi} \, \Re \{\sum_{\mu,\nu} (\hat{H_o})_{0,1}^{\mu\nu} 
G_{1\nu,0\mu}^<({\bf k},E)
\},
\end{equation}
and
\begin{eqnarray}
&&J_{\rm scatt}^{0 \rightarrow 1} = \frac{2e\rho_o}{\hbar}\int_{0}^\infty  
dE_k\int \frac{dE}{2\pi} \, 
\Re\biggr\{ \sum_\mu \Sigma_{0\mu,0\mu}^{ {\rm ret},(1)}({\bf k},E)\, 
G_{0\mu,0\mu}^{<}({\bf k},E) \, \nonumber \\
&& \qquad\qquad\qquad
 + \, \Sigma_{0\mu,0\mu}^{<,(1)}({\bf k},E)\,
G_{0\mu,0\mu}^{ {\rm adv}}({\bf k},E) \biggl\},
\end{eqnarray}
\noindent 
where $\rho_o = m_e/\pi\hbar^2$. $\Sigma^{(1)}$ is given by Eqs.
(\ref{eq.sigrough}) -- (\ref{eq.sigless}) where the sum over $n$ is restricted
to $n = 1$. 
$J_{\rm coh}$ is the coherent
contribution to the current density arising from the off-diagonal couplings
in $\hat{H}_o$, i.e., $T_1^\nu$ in $\hat{H}_{\rm SL}$ and 
$-e\mathcal{E}R_1^{\mu\nu}$ in $\hat{H}_\mathcal{E}$. 
$J_{\rm scatt}$ represents
the current contribution due to the scattering processes described
by $\hat{H}_{\rm scatt}$. We note here that the division between
coherent and scattering contributions to the current is artificial
since it depends on the basis set used in the calculation. If, for instance,
the eigenstates of $\hat{H}_{o}$ were chosen as the basis,
there would be no coherent contribution to the current.

We apply the theory to the \gaas QCL structure (Fig. 1)
described by Sirtori {\it et al.} in 
Ref. \cite{Sir98}. 
To calculate the voltage-current (V-I) characteristic, 
we evaluate the current density $J$. 
 Fig.~2 shows a plot
of voltage vs. current calculated for a lattice temperature of 77~K.
A comparison with experimental measurements (Fig. 3(a) in 
Ref. \cite{Sir98}) shows good agreement between the theoretical and
experimental results. For instance,
at a current of 4 A, the measured voltage is around 11 V, while the theoretical
curve in Fig.~2 shows a voltage of $\sim7$ V which agrees very well with
the experimental results when a residual series resistance of 
$\sim1$ $\Omega$ (see Ref. \cite{Sir98}), probably originating from the 
cladding layers, is taken into account. We comment here on the finding
of Iotti et al. \cite{Iotti}, of a strong influence of electron-electron
scattering on the I-V curve. Our calculations include interface roughness
scattering, but not electron-electron scattering. The converse is true
in the calculations of Iotti et al. where electron-electron
scattering is included, but interface roughness scattering neglected.
In both cases, I-V curves in fairly good agreement with experimental
measurements are obtained. We therefore conclude that interface roughness
and electron-electron scattering play similar roles in determining
the I-V curve, and that the I-V characteristic is not sensitive to
which type of scattering is present in the calculation.

In Fig.~3, we show the density of states $\rho_{\rm DOS} =-2 \, 
{\sum_{\nu,{\bf k}}} 
\Im [G^{\rm ret}_{0\nu,0\nu}({\bf k},E)]$
and the electron distribution function 
$f(E)= 
{\sum_{\nu,{\bf k}}} 
\Im [G^<_{0\nu,0\nu}({\bf k},E)]/ \rho_{\rm DOS}$. 
Curves at three
different applied voltages are shown: 0, 3, and 7 V.
At 0~V, we see, as expected, a thermal Fermi distribution 
corresponding to a lattice temperature of 77~K. As the applied voltage
increases, the occupation probability at higher energies increases.
In addition, phonon replicas become visible in $f(E)$.
At 7~V, the occupation
probability is higher at an energy of 0.2 eV than at lower
energies between 0 to 0.1 eV. This demonstrates the onset of   
a population inversion. At this voltage, the current is 4.5 A,
corresponding to a current density of 7.5 kAcm$^{-2}$. This
is in good agreement with the threshold current density of
7.2 kAcm$^{-2}$ for laser emission which is
reported in Ref. \cite{Sir98}.

In summary, we have applied a quantum transport theory based on nonequilibrium
Green's functions to QCL structures. The initial results for transport
properties such as current densities are in good agreement 
with experimental data reported in the literature. The calculated
electron energy distribution also demonstrates a population
inversion at a current density which agrees well with the range of
fields and currents at which lasing is observed.

\noindent
{\bf Acknowledgments:} Financial support is provided by the DFG through
FOR394.


\newpage

\begin{center}
FIGURE CAPTIONS
\end{center}

\begin{description}

\item{Fig.~1} Example of the conduction band line-up in
a QCL structure taken from Ref. \cite{Sir98}. Each period contains
16 semiconductor layers giving a period length $d = 45.3$~nm. There
are 30 periods in this structure. The in-plane area is $6 \times 10^{-4}$
cm$^{-2}$.
The Wannier functions used as
the basis set for the calculations are also shown.

\item{Fig.~2} Calculated V-I characteristic for the \gaas QCL structure
described in Ref. \cite{Sir98}. Sample area $6\times10^{-4}$ cm$^{2}$.

\item{Fig.~3} The lower row shows the distribution function of electrons for
different applied bias. The current and applied bias
are shown with each curve. The corresponding electric fields are
0, 22, and 53 kV/cm, and the current densities are 0, 85, and 7500 A/cm$^2$.
The upper row shows the corresponding density of states $\rho_{\rm DOS}$
(scaled with
 $\rho_o$) for each bias.
\end{description}

\newpage

 Fig. 1
\hskip 1truecm

\includegraphics{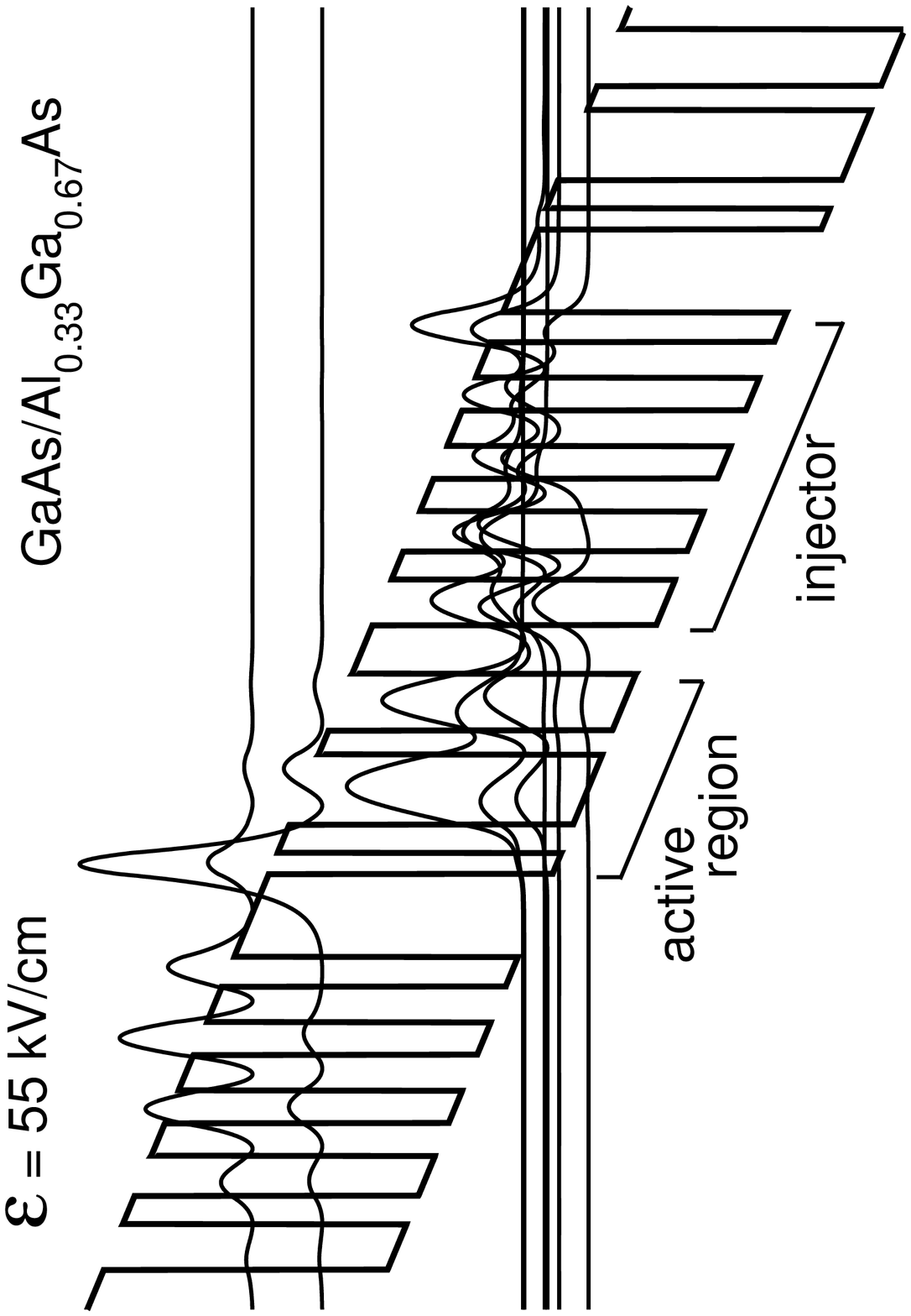}

\newpage


 Fig. 2
\hskip 1truecm

\includegraphics{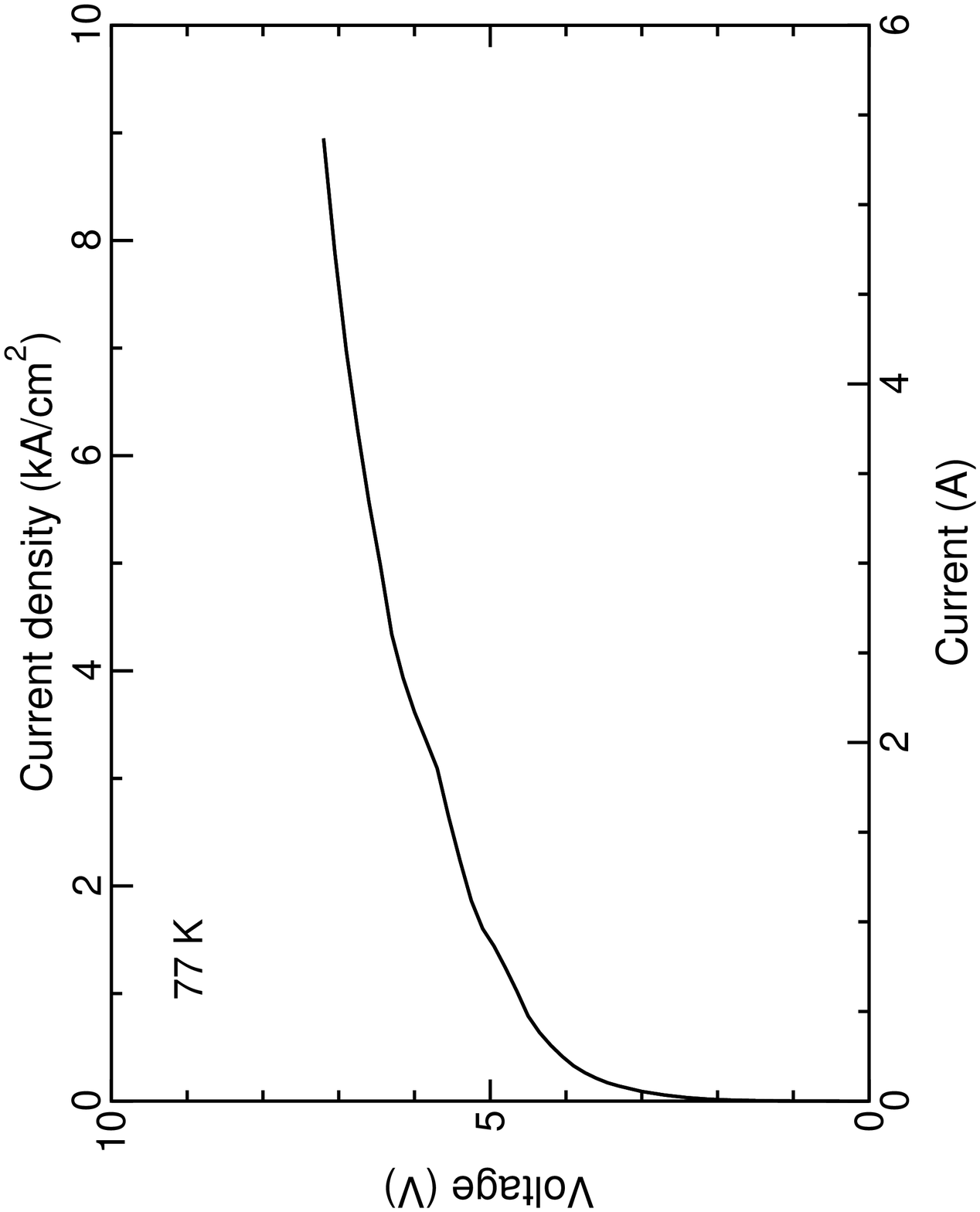}

\newpage


 Fig. 3
\hskip 1truecm

\includegraphics{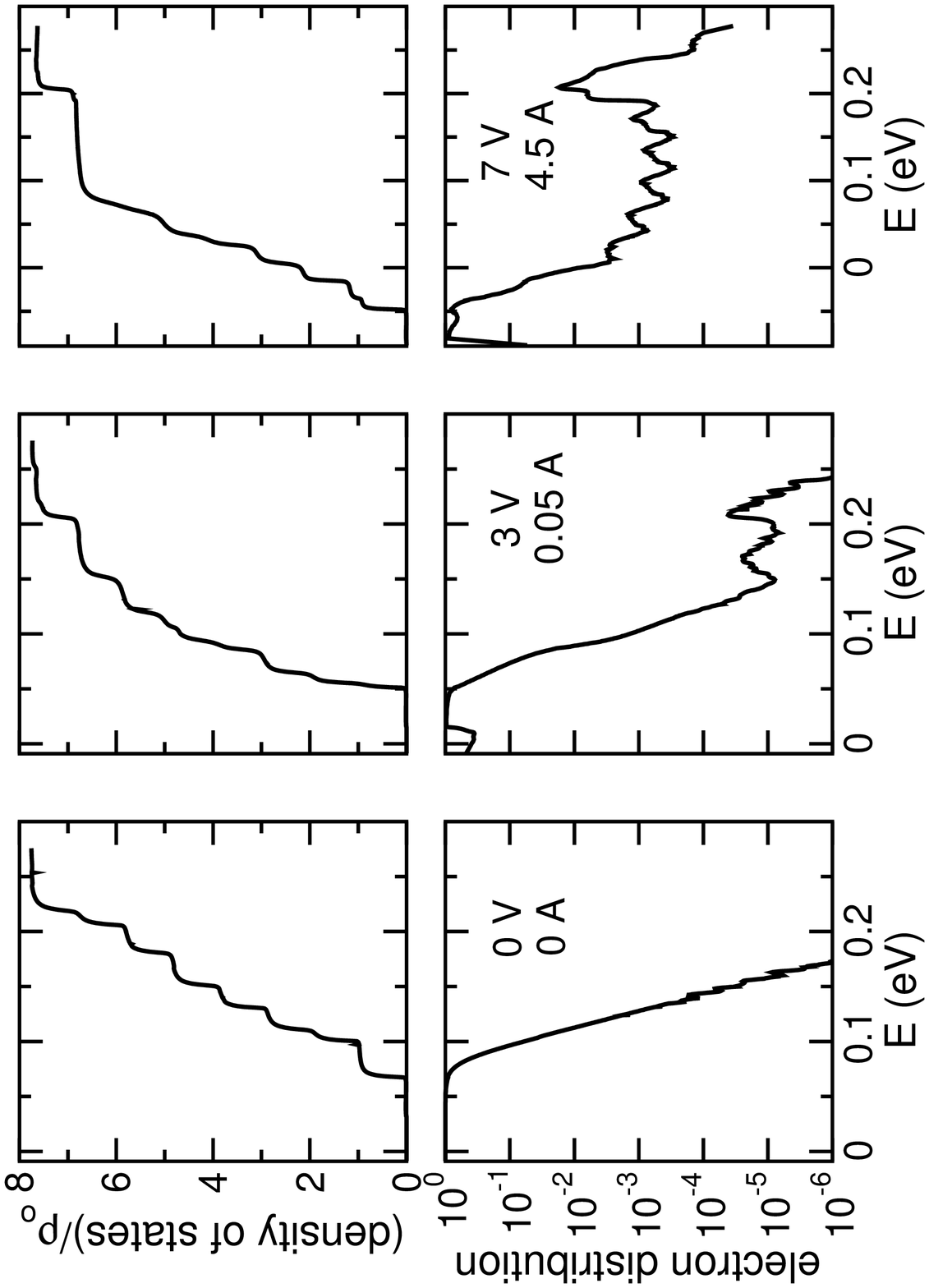}
\newpage

\end{document}